\documentclass[12pt]{article}
\usepackage[latin1]{inputenc}
\usepackage{graphicx}
\usepackage{amssymb}

\newcommand{\pa}{\partial}

\newcommand{\text}{\rm}

\newcommand{\ug}{ \; = \; }

\newcommand{\bb}{\begin{equation}}
\newcommand{\ee}{\end{equation}}
\newcommand{\bega}{\begin{eqnarray}}
\newcommand{\ega}{\end{eqnarray}}
\newcommand{\begae}{\begin{eqnarray*}}
\newcommand{\egae}{\end{eqnarray*}}

\newcommand{\h}{\hspace*{4ex}}
\newcommand{\dis}{\displaystyle}

\newcommand{\om}{\omega}

\newcommand{\cent}{\centerline}
\newcommand{\vs}{\vspace*}

\begin{document}

\baselineskip 0.8cm

\begin{center}

{\large {\bf A simple and analytical method for controlling the trajectory and multifurcation of optical beams} \
${\;^\dag }$ } \footnotetext{$^{\: ^\dag}$  Work supported by FAPESP and CNPq. \ E-mail addresses for
contacts: mzamboni@decom.fee.unicamp.br}

\end{center}

\vs{4mm}

\cent{Michel Zamboni-Rached}

\vspace*{0.2 cm}

\cent{\emph{School of Electrical and Computer Engineering, University of Campinas,}}

\cent{\emph{Campinas, São Paulo 13083-852, Brazil}}

%%\cent{{\em DECOM--FEEC, \ UNICAMP, \ Campinas, S.P., Brasil}}
%
%%%\cent{{\em University of Toronto, Toronto, ON, Canada}}

\vs{0.5 cm}

{\bf Abstract  \ --} \ In this paper it is developed a simple,
analytical and very efficient method capable to provide control of optical beam's intensity over an arbitrary curvilinear
(planar) trajectory. The same method also provides the possibility
of managing multifurcations of the optical beam. The results
presented here can have valuable applications in fields like
optical tweezers, optical lithography, atom optical guiding,
structured light, etc..

\section{Introduction}

\h The research of light beams and pulses endowed with exotic
propagation characteristics has proved to be important both, from
the academic/theoretical and applied points of view
\cite{livro1,livro2,fw1,fwbg,ahmed1}.

\h Among such exotic waves, a particular class of optical beams
has drawn the attention of many researches, namely, that of beams
with curved trajectories. Among such beams, the first one
discovered was the Airy beam \cite{airy1,airy2,airy3,airy4,airy5,airy6,airy7,airy8,airy9,ahmed2}, a solution
of the wave equation in the paraxial approximation that propagates
along a parabolic path. Soon after, non-paraxial accelerating
beams \cite{np1}, bending along circular, elliptical or parabolic
trajectories, were discovered as exact solutions of the wave
equation. In the mean time, works have been done to construct
beams with arbitrary curved trajectories \cite{curved1}; such
beams have been obtained through a phase-modulation of the optical
wavefront, creating conical bundles of rays with apexes over a
continuous focal curve chosen a priori, a method that, besides
complicated, can present limitations when considered for
non-paraxial situations.

\h In this paper, it is developed a very simple, exact and
analytical method capable to yield a three-dimensional beam whose intensity\footnote{As it is going to be clear, such control will occur beyond the intensity, being able to reach the amplitude and phase of the beam over arbitrary trajectories.} can be controlled over arbitrary (planar) curved trajectory; besides, the approach
presented here allows to manage multifurcations of the beam, which
can originate, from a single beam filament, multiple beams also
controlled at will. Finally, it is important to point out that,
although the beam trajectory is planar, the beam itself is
three-dimensional and has spot-shaped field concentration on the
planes transverse to the curved path chosen as its trajectory.

\h Due to the simplicity of the method and the wide range of
control possibilities over the optical beam, many different kind
of applications can be envisaged to it, such as in optical
micromanipulation, optical guiding of atoms, interferometry,
remote sensing, optical lithography, structured light, etc..

\section{The method}

Let us consider the following exact solution of the wave equation,
$(\nabla^2 - \pa^2_{ct})\Psi(x,y,z,t) = 0$, given by discrete
superpositions of plane waves:

\bb \begin{array}{clr} \Psi(x,y,z,t) \ug &  e^{-i\om t} \\

\

& \times \left( a^{++}\dis{\sum_{m=-M}^{M}\sum_{n=-N}^{N} A_{mn}^{++}\,e^{ik_{x m}x}e^{ik_{z n}z}e^{ik_{y mn}y}}\right.\\

\

& + \dis{\,a^{+-}\sum_{m=-M}^{M}\sum_{n=-N}^{N} A_{mn}^{+-}\,e^{ik_{x m}x}e^{ik_{z n}z}e^{-ik_{y mn}y}} \\

\

& + \dis{\,a^{-+}\sum_{m=-M}^{M}\sum_{n=-N}^{N} A_{mn}^{-+}\,e^{-ik_{x m}x}e^{ik_{z n}z}e^{ik_{y mn}y}} \\

\

& \left. + \dis{\,a^{--}\sum_{m=-M}^{M}\sum_{n=-N}^{N} A_{mn}^{--}\,e^{-ik_{x m}x}e^{ik_{z n}z}e^{-ik_{y mn}y}}\right) \,\, , \label{Psi}\end{array}\ee

with

\bb k_{y mn} \ug \sqrt{k^2 - k_{x m}^2 - k_{z n}^2} \,\,, \ee
where the coefficients $A_{mn}^{\pm \, \pm}$ are the complex
amplitudes (still unknown) of the plane waves propagating along
the directions of the wave vectors $\mathbf{k}_{mn}^{\pm \, \pm} =
\pm k_{x m}\hat{x} \pm k_{y m n}\hat{y} + k_{z n}\hat{z}$, and
$k_{x m}$, $k_{y m n}$ and $k_{z m}$ are positive and still
unknown. The $a^{\pm\,\pm}$ are constants to be fixed later.

\h The basic idea here is to use solution given by Eq.(\ref{Psi})
to construct a three-dimensional beam whose trajectory performed
by its main spot can be chosen arbitrarily as a curve $\gamma$
over the plane $(x,y=0,z)$. Besides, the method is intended to
provide a strong control over
the beam intensity along its trajectory. Finally, it is
also intended that the method enables a controlled multifurcation
of the beam, so that, for instance, from a single filamentary beam
may arise multiple beams also controlled at will.

\h To achieve such control, it is required that the method
be able to furnish an optical field whose intensity, within the region $-L_x/2 \leq x \leq L_x/2$, $0 \leq z \leq L_z$,
over the plane $y = 0$, can be chosen a priori. Mathematically, it
is demanded that

\bb |\Psi(x,y=0,z,t)|^2 \approx |F(x,z)|^2 \,\,\, \mathrm{for}\,\,
-L_x/2 \leq x \leq L_x/2,\,\, 0 \leq z \leq L_z, \,\, y = 0 \,\, ,
\label{Psiy0} \ee 

\

where $F(x,z)$, here named morphological
function, is a function chosen at will. $L_x$ and $L_z$ are
positive constants that delimit the rectangular domain (on the
plane $y=0$) where $F(x,z)$ is defined.

\h With respect to solution given by Eq.(\ref{Psi}), it is
intended that each one of the four plane wave superpositions be
able to provide the required intensity $|F(x,z)|^2$ within the
required region over the plane $y=0$. The reason for considering
four different plane wave superpositions, rather than just one
(since each one, alone, is intended to provide the desired
pattern), is due to the concern in obtaining field concentration over
any direction orthogonal to the chosen beam's trajectory, something easier to achieve when there
are both, positive and negative values to the components $ k_x $
and $ k_y $ of the wave vectors $ \mathbf{k} $ of the plane waves
making up those superpositions.

\h It is now necessary to get the values of $A_{mn}^{\pm \, \pm}$,
$k_{x m}$, $k_{z m}$ and $a^{\pm\,\pm}$ such that solution
(\ref{Psi}) be able to satisfy Eq.(\ref{Psiy0}). To do so, let
us consider the following choices

\bb k_{z n} \ug Q_z + \frac{2\pi}{L_z}n \,\, , \label{kzn} \ee

\bb k_{x m} \ug Q_x + \frac{2\pi}{L_x}m \,\, , \label{kxm} \ee

\bb A_{mn}^{p\,q} \ug \frac{1}{L_x L_z}\,
\int_{0}^{L_z}\int_{-L_x/2}^{L_x/2}\,
F(x,z)\,\exp\left(-p\,i\frac{2\pi}{L_x}m
x\right)\,\exp\left(-i\frac{2\pi}{L_z}n z\right)
\mathrm{d}x\,\mathrm{d}z \,\,, \label{Amn} \ee with $p = \pm
\,,\,\, q=\pm$ and

\bb a^{++} \ug a^{+-} \ug a^{-+} \ug \frac{1}{2}\;;\,\,\, a^{--} \ug - \frac{1}{2} \,\,. \label{apm}  \ee

\h In the above equations, $Q_z$ is a constant chosen to define
the paraxiality degree of the resulting beam; once it is chosen,
the choice of $Q_x$ regulates the average inclination of the plane
waves, which constitute the resulting beam, relative to the $x$
direction and, therefore, also defines the average inclination
with respect to the $y$ direction. Such choices have influence
over the degree of field concentration on the planes transverse to
the curved path chosen as the beam's trajectory. The coefficients
$A_{mn}^{\pm\,\pm}$ are the Fourier coefficients of the
morphological function $F(x,z)$, and it is simple to verify that

\bb A_{mn}^{++} \ug A_{mn}^{+-} \ug A_{-mn}^{-+} \ug A_{-mn}^{--}  \label{AA} \ee

\h The reason for choosing the values of $a^{\pm\,\pm}$ as given by Eq.(\ref{apm}) will be clear soon.

\h With the choices given by Eqs.(\ref{kzn},\ref{kxm},\ref{apm}),
the main solution (\ref{Psi}) acquires the following form over the
plane $y=0$:

\bb \begin{array}{clr} \Psi(x,y=0,z,t) \ug & \frac{1}{2}\,e^{-i\om t}e^{i Q_z z} \\

\

& \times \left[e^{i Q_x x}\dis{\sum_{m=-M}^{M}\sum_{n=-N}^{N} A_{mn}^{++}\,\exp\left(i\frac{2\pi}{L_x}m x\right)\,\exp\left(i\frac{2\pi}{L_z}n z\right)}\right.\\

\

& + \,\dis{\,e^{i Q_x x}\sum_{m=-M}^{M}\sum_{n=-N}^{N} A_{mn}^{+-}\,\exp\left(i\frac{2\pi}{L_x}m x\right)\,\exp\left(i\frac{2\pi}{L_z}n z\right)} \\

\

& + \,\dis{\,e^{-i Q_x x}\sum_{m=-M}^{M}\sum_{n=-N}^{N} A_{mn}^{-+}\,\exp\left(-i\frac{2\pi}{L_x}m x\right)\,\exp\left(i\frac{2\pi}{L_z}n z\right)} \\

\

& \left. - \, \dis{\,e^{-i Q_x x}\sum_{m=-M}^{M}\sum_{n=-N}^{N} A_{mn}^{--}\,\exp\left(-i\frac{2\pi}{L_x}m x\right)\,\exp\left(i\frac{2\pi}{L_z}n z\right)}\right] \,\,. \label{Psiy01}\end{array}\ee

Now, due to Eqs.(\ref{Amn},\ref{AA}), all double summations in
Eq.(\ref{Psiy01}) are equal, resulting, approximately, in
$F(x,z)$. With this, Eq.(\ref{Psiy01}) can be written as

\bb \begin{array}{clr} \Psi(x,y=0,z,t) & \approx \frac{1}{2}\,e^{-i\om t}e^{i Q_z z}\left(e^{i Q_x x}+e^{i Q_x x}+e^{-i Q_x x}-e^{-i Q_x x}\right)F(x,z) \\

\\

 & \ug  e^{-i\om t}e^{i Q_z z}e^{i Q_x x}F(x,z) \,\, , \label{Psiy02} \end{array}\ee

 within the region $-L_x \leq x \leq L_x$, $0 \leq z \leq L_z$, over the plane $y=0$. Therefore,
the required condition (\ref{Psiy0}) over $\Psi(x,y,z,t)$ is
naturally achieved.

\h At this point, it is worth noting three points:

\begin{itemize}
    \item {Equation (\ref{Psiy02}), the main consequence
of the method, not only confirms that condition (\ref{Psiy0}) is
satisfied but goes further, showing that amplitude and phase could be also be chosen \emph{a priori.}}
    \item {one could choose as main solution
just one of the double summations of Eq.(\ref{Psi}) that, even
then, the desired result given by Eq.(\ref{Psiy0}) would be
achieved. However, such a procedure has been avoided as it could
result in excessive energy flux along a given direction; for
instance: if we chose only the first summation, there would be
unnecessary amount of energy flow along directions close to that
given by $\mathbf{k} = Q_x\hat{x} + \sqrt{k^2 -
Q_x^2-Q_z^2}\,\hat{y} + Q_z \hat{z} $, something that could affect
the field concentration on the planes transverse to the curved
path chosen to the resulting beam.}
    \item {In this paper, the main goal is to get strong control over optical beams, however, it should be clear that the
    present method also can be used to create very intricate patterns, like images, on the plane
    $(x,y=0,z)$.}
\end{itemize}

\h Naturally, since the method proposed here is based on discrete
superpositions of plane waves, the pattern chosen to the beam has
to be considered only within a given spatial volume, being
periodic outside it. However, this fact should not be seen as a
limitation, since it can be easily overcome by an appropriate
apodization, something that will be addressed in an upcoming work.

\h As we are going to show in the following section, the method
presented above is capable to furnish wave beams with unprecedent
spatial configurations.

%\newpage

\section{Applying the method}

\h The method presented in the previous section showed how to get
a strong control over an optical beam. It is just necessary to
choose the morphological function, $F(x,z)$, and then use
Eqs.(\ref{Psi},\ref{kzn},\ref{kxm},\ref{Amn},\ref{apm}) to model
the wave field according to it. The main interest of this paper is
to use such method to get arbitrary curved beams and also control their intensities along their curved paths.
Due to this,
it is worth to describe how to compose the morphological function
once the beam's trajectory and the beam's intensity
along it have been chosen.

\h Suppose that the chosen beam's trajectory, $\gamma$, can be
written, within the domain $0 \leq z \leq L_z$, $-L_x/2 \leq x
\leq L_x/2$ and $y=0$, by a functional relation of the type

\bb x = f(z) \,\,, \label{f} \ee more rigorously

\bb \gamma \mathrm{:} \mathbb{R} \rightarrow \mathbb{R} \, ,
\,\,\, \gamma (z) = (f(z),z) \,\, . \ee

%\bb \gamma \mathrm{:} \mathds{R} \rightarrow \mathds{R} \, , \,\,\, \gamma (z) =
%(f(z),z) \,\, . \ee

A very simple way to get a morphological function $F(x,z)$ that
ensures the beam's trajectory as given by function (\ref{f}) is by
writing:

\bb F(x,z) \ug \exp\left(-q^2\left[x - f(z)\right]^2\right) \,\,,
\label{F1}\ee

where the parameter $q^2>0$ defines the field
concentration degree over the path $\gamma$, and can be connected
to the spot size of the beam along the $x$ direction through
$\Delta x = \sqrt{2}/q$. Naturally, it is possible to choose many
other morphological functions, besides that one given by
Eq.(\ref{F1}), which ensure the field concentration along the
curve $\gamma$ given by Eq.(\ref{f}).

\h Now, the morphological function (\ref{F1}) can be improved to
incorporate the control over the beam's intensity (actually, over its amplitude and phase) along
its trajectory. This can be done in a simple way by writing

\bb F(x,z) \ug \Theta(z)\,\exp\left(-q^2\left[x -
f(z)\right]^2\right) \,\,. \label{F}\ee

\h In Eq.(\ref{F}), the exponential function guarantees the beam's
trajectory, $\gamma$, and the complex function $\Theta(z)$ can be
used to incorporate the desired intensity to the beam
along $\gamma$.

\h Once having $F(x,z)$, the desired (and exact) beam solution is
given by Eqs.(\ref{Psi},\ref{kzn},\ref{kxm},\ref{Amn},\ref{apm}),
where the values of $Q_x$ and $Q_z$ are chosen according to the
desired degree of paraxiality and field concentration along directions orthogonal to the chosen beam's trajectory

\h It is important to note that the morphological function given
by Eq.(\ref{F}) will be useful only for cases with absence of
multifurcations, i.e., only in cases where $f(z)$ is a
function (for a given $z$, there is just one value to $f$
within the domain considered). In the case of multifurcations, the
morphological function can be given by a summation of functions of
the type (\ref{F}), as we are going to see in the second example.

\h Now, the method is applied for obtaining some interesting
diffraction resistant beams with curved trajectories. In all of the following examples, the coefficients $A_{mn}^{\pm \pm}$, given by Eq.(\ref{Amn}), are calculated numerically from the morphological function, $F(x,z)$, in question. Of course, the main solution (\ref{Psi}) remains being analytical and exact.

\h Here, it will be considered $\lambda = 0.532 \mu$m, widely used
in many applications.

\

\textbf{First Example:}

\h In this first example, it is desired a beam whose spot of
constant intensity initially performs a sinusoidal trajectory,
which becomes rectilinear and that, after some distance, becomes
sinusoidal again. Such path can be represented by the following
function:

\bb x \ug f(z) \ug a\sin\left(\eta
z\right)\left[1-\exp\left(-s^{10}\left(z -
z_0\right)^{10}\right)\right] \,\,, \label{f1} \ee

which, by setting $a=4\times10^{-4}$, $\eta=125.66\,\mathrm{m}^{-1}$, $s=13.3\,\mathrm{m}^{-1}$ and $z_0=15$cm, yields the path shown in Fig.1 and adopted in this example.

\h Once the trajectory was chosen, the morphological function,
$F(x,z)$, is obtained by inserting $f(z)$, given by Eq.(\ref{f1}),
into Eq.(\ref{F}), where it is considered $\Theta(z)=1$. For the
beam, it is desired an intensity spot width (along the $x$
direction) $\Delta x = 0.1$mm, which demands $q=1.41\times
10^{4}\,\mathrm{m}^{-1}$ in Eq.(\ref{F}).

\h With this, the resulting beam is obtained through
Eqs.(\ref{Psi},\ref{kzn},\ref{kxm},\ref{Amn},\ref{apm}), where it
was chosen $L_x=4$mm, $L_z=30$cm, $Q_z=0.99985 k$, $Q_x =
\sqrt{k^2 - Q_z^2}/\sqrt{2}$ and $M=N=20$. Figure 2 shows the
resulting beam intensity on the plane $y=0$, i.e,
$|\Psi(x,y=0,z,t)|^2$ and it is very clear that the beam's
trajectory is precisely the one chosen.

\begin{figure}[!h]
\begin{center}
 \scalebox{.65}{\includegraphics{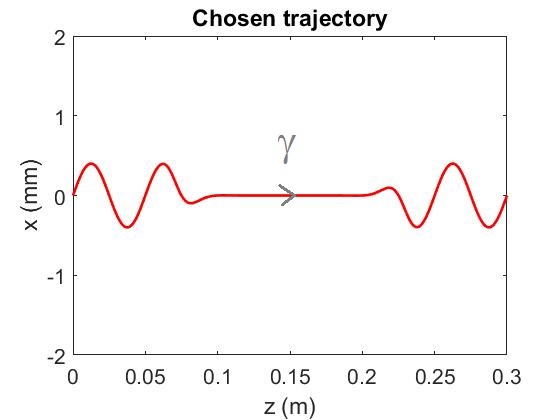}}
\end{center}
\caption{Trajectory chosen for the beam in the first example.} \label{fig1}
\end{figure}

\begin{figure}[!h]
\begin{center}
 \scalebox{.7}{\includegraphics{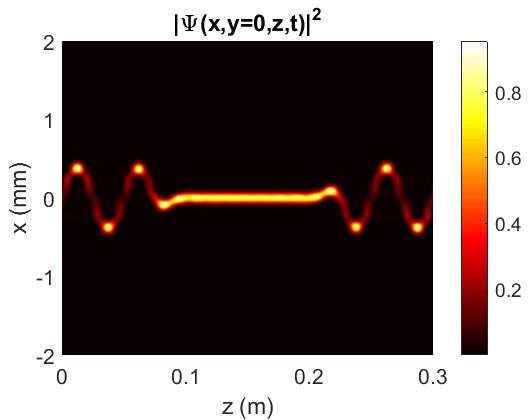}}
\end{center}
\caption{The intensity of the resulting beam on the plane $y=0$, i.e,
$|\Psi(x,y=0,z,t)|^2$. It is very clear that the beam's
trajectory is precisely the one chosen in Fig.1.} \label{fig2}
\end{figure}

%\newpage

Figure 3 depicts the 3D intensity of the resulting beam, from where one can see it is transversely concentrated 
along the beam's trajectory.

\begin{figure}[!h]
\begin{center}
 \scalebox{.48}{\includegraphics{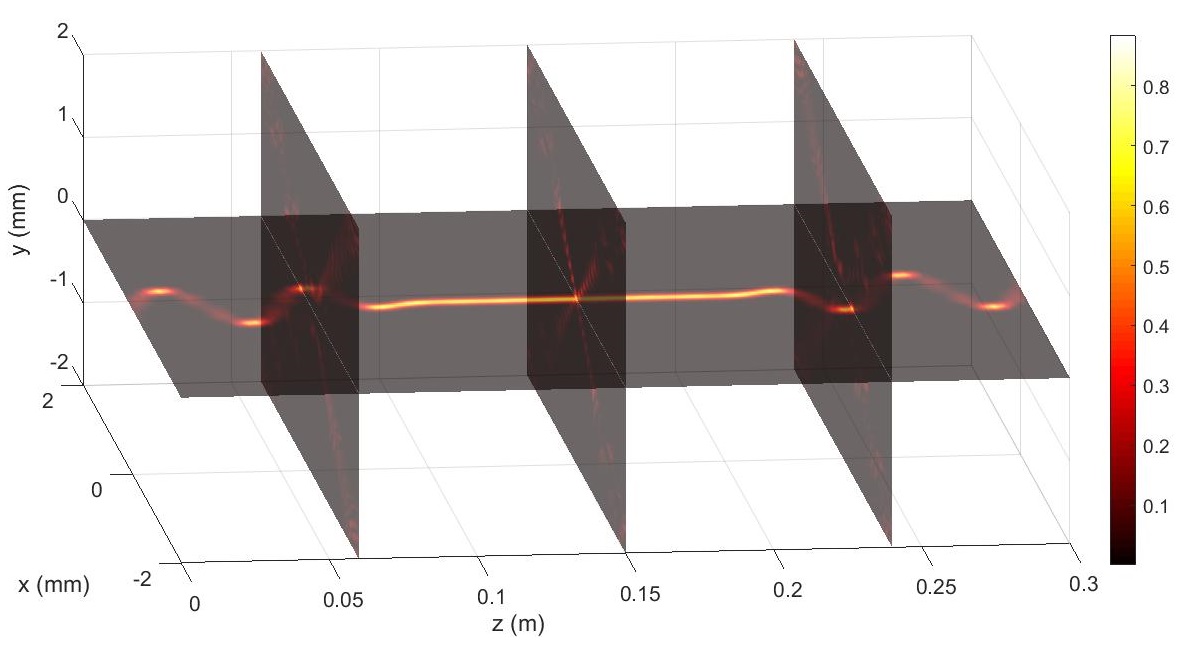}}
\end{center}
\caption{The 3D intensity of the resulting beam of the first example. One can see it is transversely concentrated
along its trajectory.} \label{fig3}
\end{figure}

\h It should be clear that the energy flux occurs mainly in the positive $z$ direction.

%%\newpage

\

\textbf{Second Example:}

\h In this first example, it is desired a beam performing a finite
sinusoidal trajectory, i.e., the beam only has constant and
non-negligible intensity over a finite range of the sinusoidal
trajectory.

\h The sinusoidal path can be represented by the following
function $f(z)$:

\bb x \ug f(z) \ug a\sin\left(\eta z\right) \,\,, \label{f2} \ee

which, by setting $a= 1.67\times10^{-4}$ and $\eta=125.6 \,\mathrm{m}^{-1}$, yield the path shown in
Fig.4 and adopted in this example. In that figure, the continuous
line shows the part ($0.03\,\mathrm{m}\leq z \leq
0.15\,\mathrm{m}$) of the trajectory over which a constant
intensity is desired and the doted line shows the rest of the
trajectory, where it is desired a negligible (ideally zero)
intensity.

\begin{figure}[!h]
\begin{center}
 \scalebox{.65}{\includegraphics{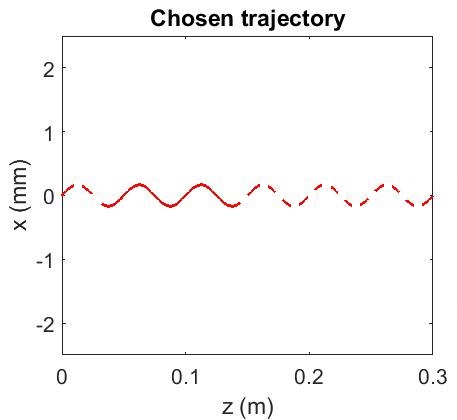}}
\end{center}
\caption{Trajectory chosen for the beam in the second example.} \label{fig1}
\end{figure}

\h With the trajectory equation at hand, the morphological
function, $F(x,z)$, is obtained by inserting $f(z)$, given by
Eq.(\ref{f2}), into Eq.(\ref{F}). Here, it is desired a intensity
spot width (along the $x$ direction) $\Delta x = 0.125$mm, which
demands $q=1.31\times 10^{4}\,\mathrm{m}^{-1}$ in Eq.(\ref{F}).
Also, since it is desired a non-negligible intensity just for
$0.03\,\mathrm{m}\leq z \leq 0.15\,\mathrm{m}$, it is chosen

\bb \Theta(z)= \exp\left(-s^{10}\left(z - z_0\right)^{10}\right)\,\,,\label{teta2}\ee

with $s= 16.67 \,\mathrm{m}^{-1}$ and $z_0 = 0.09$m. With this, the morphological function, $F(x,z)$,
is obtained by inserting Eqs.(\ref{f2},\ref{teta2}) into Eq.(\ref{F}).

\h Now, the resulting beam is obtained through
Eqs.(\ref{Psi},\ref{kzn},\ref{kxm},\ref{Amn},\ref{apm}), where it
was chosen $L_x=5$mm, $L_z=30$cm, $Q_z=0.9998 k$, $Q_x =
\sqrt{k^2 - Qz^2}/\sqrt{2}$ and $M=N=29$. Figure 5 shows the
resulting beam intensity on the plane $y=0$, i.e,
$|\Psi(x,y=0,z,t)|^2$ and it is very clear that the beam's
trajectory is precisely the one chosen.

Figure 6 depicts the 3D intensity of the resulting beam, where one can see it is transversely concentrated along its trajectory.

\h It should be clear that the energy flux occurs mainly in the
positive $z$ direction.

\begin{figure}[!h]
\begin{center}
 \scalebox{.75}{\includegraphics{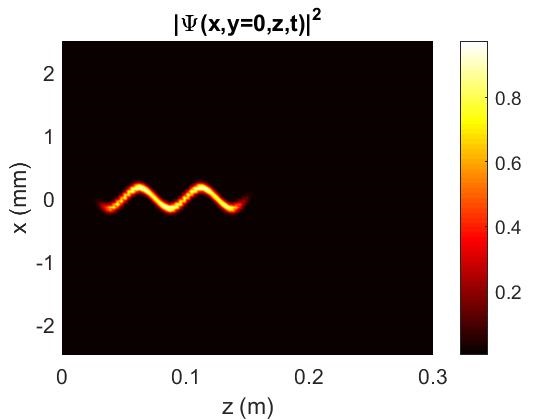}}
\end{center}
\caption{The intensity of the resulting beam of the second example on the plane $y=0$, i.e,
$|\Psi(x,y=0,z,t)|^2$. It is very clear that the beam's
trajectory is precisely the one chosen in Fig.4.} \label{fig2}
\end{figure}

\begin{figure}[!h]
\begin{center}
 \scalebox{.5}{\includegraphics{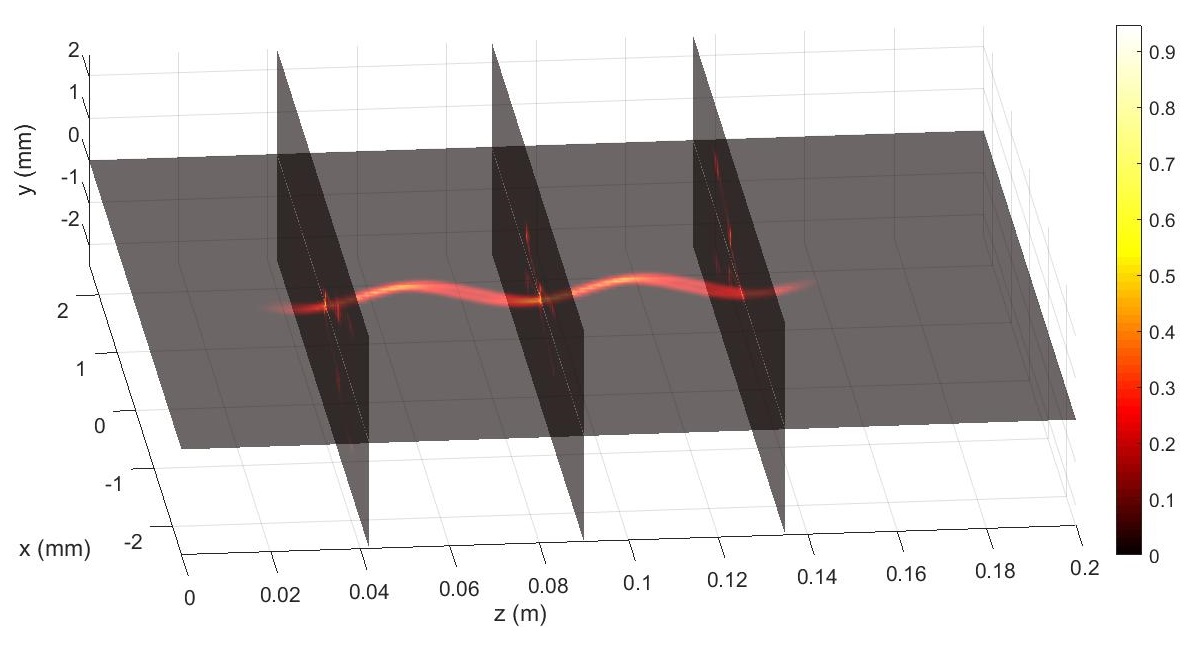}}
\end{center}
\caption{The 3D intensity of the resulting beam of the second example. One can see it is transversely concentrated
along its trajectory.} \label{fig3}
\end{figure}

\newpage

%\newpage

\textbf{Third Example:}

\h This example will deal with a beam that, at given point,
suffers a multifurcation in three beams, two of them rectilinear
and one with a sinusoidal pattern. After a certain distance, the
three beams merge into just one (again) and, in the whole process,
the intensity remains constant.

\h Of course, it is not possible to represent such a situation
with only one path function $ f (z) $. Actually, that
multifurcation will be represented by three path functions, $ f_1
(z) $, $ f_2 (z) $ and $ f_3 (z) $, which will be used to compose
a single morphological function $F(x,z) $. In this example, each of the
paths is written as:

\bb \begin{array}{clr} x = f_1(z) = a_1\exp\left(-s^4(z-z_0)^4\right) \\

\\

x = f_2(z) \ug a_2\cos(\eta z)\exp\left(-s^4(z-z_0)^4\right) \\

\\

x = f_3(z) = a_3\exp\left(-s^4(z-z_0)^4\right) \,\, , \end{array} \ee

with the following values considered for the
parameters: $a_1= 4\times
10^{-4}\,\mathrm{m}^{-1}$, $a_2 = 1\times 10^{-4}\,\mathrm{m}^{-1}
$, $a_3 = -4\times 10^{-4}\,\mathrm{m}^{-1}$,
$\eta=125.66\,\mathrm{m}^{-1}$, $s=16.66\,\mathrm{m}^{-1}$ and
$z_0=15$cm.

Figure 7 shows the composition of the 3 paths defined by the $f_1(z)$, $f_2(z)$ and $f_3(z)$.

\begin{figure}[!h]
\begin{center}
 \scalebox{1}{\includegraphics{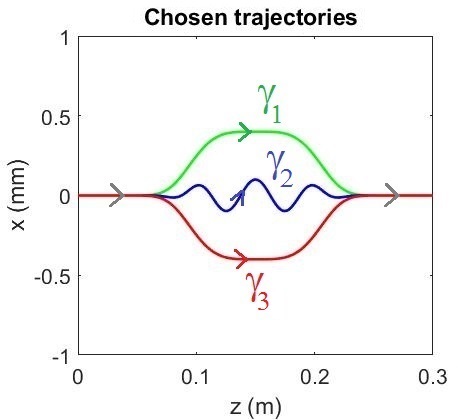}}
\end{center}
\caption{Composition of the 3 paths chosen to represent a beam multifurcation in three beams which, after a certain distance, merge into just one again.} \label{fig7}
\end{figure}

\h Now, concerning the morphological function $F(x,z)$ to this
case, as it was said before, it can't be of the form given by Eq.(\ref{F}). However, it is possible to construct a single
morphological function capable of embracing all trajectories by
making

\bb \begin{array}{clr} F(x,z)  & =  \Theta_1(z)\left(\exp\left(-q^2\left[x -
f_1(z)\right]^2\right)\right) \\

\\

 & + \,\,\, \Theta_2(z)\left(\exp\left(-q^2\left[x -
f_2(z)\right]^2\right)\right) \\

\\

 & + \,\,\, \Theta_3(z)\left(\exp\left(-q^2\left[x -
f_3(z)\right]^2\right)\right) \,\,, \end{array} \label{F2}\ee

where $\Theta_1(z)$, $\Theta_2(z)$ and $\Theta_3(z)$ are functions
that can be used to control the amplitude and phase of the
resulting beam along the multifurcated paths. In this example, it
is demanded a constant beam intensity along all trajectories, which can be achieved by making $\Theta_1(z) = \Theta_2(z) =
\Theta_2(z) = 1+2\exp(-s^4(z-z_0)^4)$, so resulting in the
following final morphological function:

\bb \begin{array}{clr} F(x,z)  & =  \left[1+2\exp\left(-s^4\left[z-z_0\right]^4\right)\right]\,\left[\exp\left(-q^2\left[x -
f_1(z)\right]^2\right)\right. \\

\\

 & + \,\,\, \exp\left(-q^2\left[x -
f_2(z)\right]^2\right) \\

\\

 & + \,\,\, \left.\exp\left(-q^2\left[x -
f_3(z)\right]^2\right)\right] \,\,, \end{array} \label{F22}\ee

where we choose $q = 2.83 \times 10^{4}\,\mathrm{m}^{-1}$ for obtaining a spot width (along the $x$ direction) $\Delta x = 50\mu \mathrm{m}$.

\h With this, the resulting beam is obtained through
Eqs.(\ref{Psi},\ref{kzn},\ref{kxm},\ref{Amn},\ref{apm}), where it
was chosen $L_x=2$mm, $L_z=30$cm, $Q_z=0.99991 k$, $Q_x =
\sqrt{k^2 - Qz^2}/(2\sqrt{2})$ and $M=N=20$. Figure 8 shows the
resulting beam intensity on the plane $y=0$, i.e,
$|\Psi(x,y=0,z,t)|^2$. It is very clear that the beam has a
multifurcation, splitting in three beams (as required) and
evolving to a single beam again.

\begin{figure}[!h]
\begin{center}
 \scalebox{.6}{\includegraphics{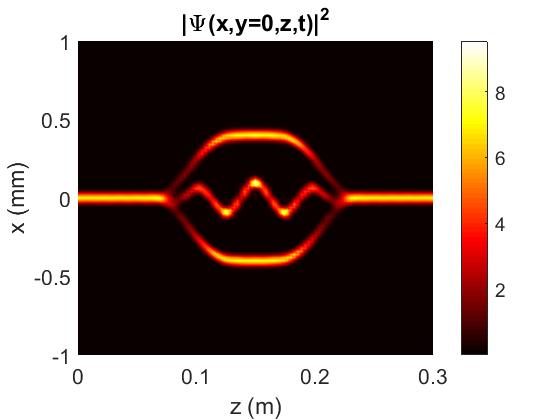}}
\end{center}
\caption{The intensity of the resulting beam of the third example on the plane $y=0$, i.e,
$|\Psi(x,y=0,z,t)|^2$. It is very clear that the beam's multifurcation behavior is precisely the one chosen in Fig.7.} \label{fig8}
\end{figure}

Figure 9 depicts the 3D intensity of the resulting beam, where it
is possible to see that the field is transversely concentrated along the trajectories.

\h Again, it should be clear that the energy flux occurs mainly in the positive $z$ direction.

%\newpage

\begin{figure}[!h]
\begin{center}
 \scalebox{.48}{\includegraphics{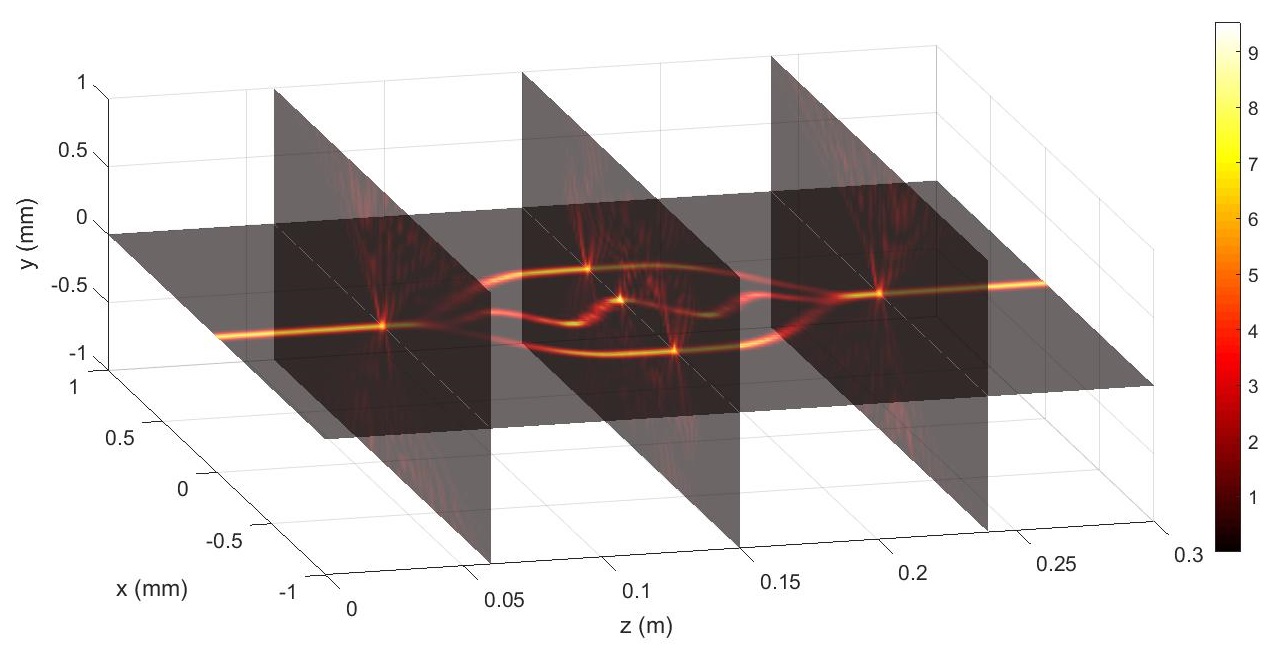}}
\end{center}
\caption{The 3D intensity of the resulting (multifurcated) beam of the third example. One can see it is transversely concentrated
along the trajectories.} \label{fig9}
\end{figure}

\section{Conclusions}

In this paper, it was developed a very simple, exact and
analytical method capable to yield a three-dimensional beam whose
intensity can be controlled over arbitrary curved trajectories. The same method also allows to manage
multifurcations of the beam, which can originate, from a single
beam filament, multiple beams also controlled at will. Finally, it
is important to point out that, although the beam trajectory is
planar, the beam itself is three-dimensional and has spot-shaped
field concentration on the planes transverse to the curved path
chosen for the beam's trajectory.

\h The method can have many different kind of applications, such as in optical micromanipulation, optical guiding of atoms, interferometry, remote sensing, optical lithography, structured light, etc..

\section*{Acknowledgements}
Thanks are due to partial support from FAPESP (under grant 2015/26444-8) and from CNPq
(under grant 304718/2016-5).

\end{document}